\documentstyle[11pt]{article}
\input epsf

\begin{document}

\title{\bf Periodic One-Dimensional Hopping \\
model with one Mobile \\ Directional Impurity\\}

\author{  \\ Z. Toroczkai and R.K.P. Zia\\
  \\
\em Department of Physics and
Center for Stochastic Processes \\ 
\em in Science and
Engineering, Virginia Polytechnic  
Institute \\ 
\em and State University,
Blacksburg, V.A. 24061-0435, USA}

\maketitle

\vspace*{1.0cm}

\thispagestyle{empty} 

\begin{abstract}
Analytic solution is given in the steady
state limit $t\to \infty $ for the
system of Master equations describing
a random walk on one-dimensional
periodic lattices with arbitrary
hopping rates containing one mobile,
directional impurity (defect bond).
Due to the defect, translational
invariance is broken, even if all other
rates are identical. The structure
of Master equations lead naturally to
the introduction of a new entity,
associated with the walker-impurity
pair which we 
call the quasi-walker. The
velocities and diffusion constants for
both the random walker and impurity
are given, being simply related to
that of the quasi-particle through
physically meaningful equations.
Applications in driven diffusive systems
are shown, and connections with the
Duke-Rubinstein reptation models for gel
electrophoresis are discussed.
\end{abstract}
\vspace*{1.0cm}
\hspace*{1cm}PACS: $02.50.-r$, $05.40.+j$,
$05.60.+w$, $83.10.Nn$, $83.10.Pp$ \\

\vspace*{0.5cm}

\hspace*{6cm}$\overline
{\hspace*{2cm}\mbox{\small \em 
TO APPEAR IN J.STAT.PHYS}}$

\newpage
 
\setcounter{page}{1}

\setlength{\topmargin}{-0.5cm}

\section{Introduction}

In equilibrium statistical mechanics the
number of exactly (analyticaly)
solvable models is very limited
(for a review see \cite{Bax}). In
nonequilibrium statistical mechanics
this number is considerably smaller,
and no review has been published,
to our best knowledge, so far. Part of
this difference lies in the starting
points. In equilibrium physics, one
usually begins with the well-known
Boltzmann distribution. By contrast,
arriving at this step is already
non-trivial, even for steady states, since
one must solve, say, a Master
equation first. To find an analytical
solution for such equations, for
typical nonequilibrium systems, is usually
impossible because of the very large
number of different equations involved,
since the unknown is a function
of the {\em configurations }of the system.
However, we can hope to find
analytic solutions if the dimensionality of our
system is low enough. As an
example, consider the random walk
of a single particle on a
one dimensional (1D) lattice,
with time-independent hopping
rates $W_{i,j}$, where $W_{i,j}$
represents the probability for the
particle to jump from site $j$ to
site $i$ in one step (or per unit time).
These rates may be arbitrary, so that they
are not required to satisfy the condition
for detailed balance, so as to describe
a general, nonequilibrium
system. With the additional assumptions
(a) that only jumps to the nearest
sites are allowed ($\mid \!i-j\!\mid =1$)
and (b) that periodic boundary
condition (PBC) is imposed, this model
was analyticaly solved in the steady
state limit by Derrida in 1983 (see
\cite{Derrida}). Explicit expressions
for the steady state distribution,
velocity and diffusion constant were
given. However, if, say, jumps to next
nearest neighbours are allowed, an
analytical solution seems to be
impossible to give \cite{Derrida}. On the
other hand, since 1983, several
other types of 1D systems (e.g., \cite
{asepobc}, \cite{BeaRoy}) have
been solved.

In the next two subsections of the
Introduction we explicitely present on a
``checkerboard'' the one-dimensional
periodic hopping model without and with
a mobile directional impurity.
Then we give a short description
of the main body of the paper.

\subsection{One-dimensional hopping
model without impurities}

The hopping model presented by
Derrida in \cite{Derrida} can best be
visualized as a $(N+1)\times 1$
checkerboard, i.e., a string of $N+1$
squares).
This is filled with $N$ pieces (particles)
numbered from $1$ to $N$
(from left to right) and one hole,
which will play
the role of the random walker
(Fig.1) . The periodic boundary
condition is equivalent to identifying
the squares at the two ends as nearest
neighbours. By saying ``that the
hole is at site $n$'' we mean that the
empty square is found in
between particles with labels
$n$ and $n+1$.
Since the dynamics will be restricted to
particle-hole exchange only, we may
regard the hole as the random walker,
and will use the terms`hole' and
`walker' interchangibly. The proposition
``the hole jumps from site $n$ to
site $n+1$ in the next time step''
on Fig.1 is equivalent to ``shifting the
hole from the square in between
particles $n$ and $n+1$ to the square in
between particles $n+1$ and $n+2$''.
The rules of evolution are as follows:\\
1) a square can be at most occupied
by a single particle,\\
2) a particle can interchange
squares only with the empty
square (hole) and
only if that is a nearest neighbour, \\

\begin{figure}
\vspace*{-4cm}
\hspace*{2.2cm}\epsfxsize = 4 
in \epsfbox{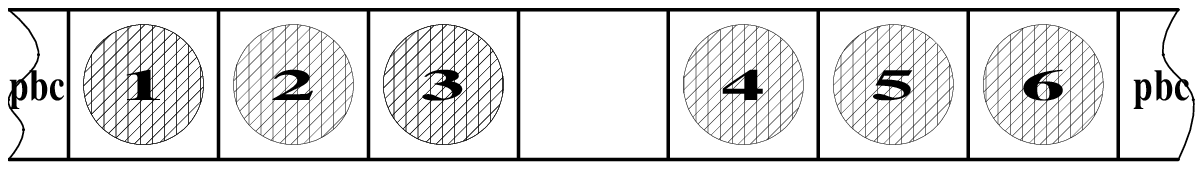}
\vspace*{-4cm}
\caption{The translationally invariant 
one dimensional hopping model with 
periodic boundary conditions, on a checkerboard.
The particles are represented by numbered
circles (pieces).}
\end{figure}
\noindent
3) the particles are not allowed to
interchange positions among themselves, \\
4) whenever the hole at site $n$
jumps to the left (to site $n-1$),
it does with probability $W_{n-1,n}$,
and, for a jump to the right 
(to site $n+1$), the
probability involved is $W_{n+1,n}$.

Note that the system above is {\em
translationally invariant}, in the sense
that the jump rate of any particular
particle-hole pair is independent of
its location on the checkerboard.
In particular, after the hole jumps
through the entire system a few times,
(effectuating complete revolutions)
the string of particles will be
displaced, relative to the ``board''.
Nevertheless, the physics of the
evolution remains unaltered. Next, we
break this invariance by introducing a
``kink'' in the checkerboard.

\subsection{The model with one
mobile directional impurity}

In this paper we introduce a mobile,
directional impurity into the periodic
hopping model above. In terms of the
checkerboard, this is a specific
``bond'' between two adjacent squares, such
that rates of particle-hole exchanges
across it are different
(in the sense that they differ  from
the $W$ rates) . Thus, we
also refer to it as a ``defect bond''
(DB). Since the string of particles
can be displaced relative to the
checkerboard, this DB {\em can move},
relative to the string. In this sense,
we use the term ``mobile directional
impurity''. A more transparent way to
display the model is represented in
Fig.2, where the DB is shown as a kink
in the checkerboard. Only across this
kink are the probabilities of the
particle-hole exchange different from
the $W$'s specified above. The
motivation for such a kink will be
discussed below. Here, we simply state
the model.
In addition to the rules 1)-4) from
the previous subsection, we have one
more, specifying the jump rates through
the kink:\\ 5) the probabilities for
the interchange of the hole with a
particle through the DB (thick dashed line) are
$q$, if the hole moves upward, and
$q^{\prime }$, if downward, {\em
independent} of the particle involved.
This is a simplifying assumption,
allowing less involved formulae. Our
methods will lead to an analytic
solution even if these probabilities do
depend on the particle.

Note that, in both models, the order of
the particles is {\em never}
changed, i.e., there is no mixing among
the particles. Therefore, when the
hole is in between particles $n$ and
$n+1$, the hopping probability to the
left (right) is always $W_{n-1,n}$
($W_{n+1,n}$), except when the DB is
involved. However, the string of
particles as a {\em whole} shifts with
respect to the checkerboard (and the DB),
as the hole wanders. 

\begin{figure}
\vspace*{-4cm}
\hspace*{2cm}\epsfxsize = 4 in \epsfbox{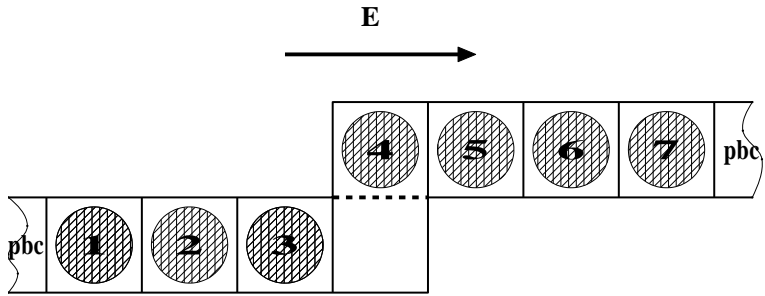}
\vspace*{-3cm}
\caption{Translational invariance is broken
by introducing a kink in the checkerboard
(lattice).The jumps across the kink (dashed line)
obey different rates, i.e., $q$ and $q'$,
independently of the particle involved. If the
particles carry charges, then in the case of
an external electric field {\bf E} the jumps
across the kink are {\em not} biased.}
\end{figure}
\noindent
{\em Fixing
our reference frame to the string
of particles}, which turns out to be
mathematically convenient, we see
that the DB `moves', if and only if the
hole passes through it. Further, the
jump rates from site $n$ to $n\pm 1$
are not always $W_{n\pm 1,n}$, but take
on the value $q$ or $q^{\prime }$
whenever the jump is across the DB.
Naively, one may expect that an analytic
solution is impossible, given the
seemingly time-dependent rates.
Nevertheless, we are able to find a
solution. The key observation is the
existence of a specific linear
combination of the positions of the
hole and the DB, with which the system
of Master equations can be rewritten
in a form describing the hopping of a
{\em single} random walker (called
quasi-walker)
on a chain of length $N(N+1)$.
Exploiting the results of Derrida, the
probability distribution, velocity
and diffusion constants of the
quasi-walker, as well as those for
the hole and the DB, can be explicitly
computed.

An immediate application for the model
with directional impurity is a
generalized asymmetric exclusion process
(ASEP) \cite{Derrida}, \cite{asepobc}.
Consider the pieces as charged
particles. More precisely, let the
particles be charged with $Q_n$ and
let us apply an external electric field
${\bf E}$ pointing along the chain
(See Fig.2). Neglecting the interaction
among charges, the hopping rates are
solely determined by the influence of
${\bf E}$ on the charges. Away from
the kink, these rates would be biased,
due to ${\bf E}$. However, across the
kink (or DB), particle hops are {\em
perpendicular} to the external field,
so that the corresponding
probabilities ($q$ and $q^{\prime }$)
are {\em not biased} by ${\bf E}$.

Another application, which is also the
physical motivation of our model,
comes from the connection to the
reptation model of Duke-Rubinstein for
gel electrophoresis \cite{rubi}.
In Ref. \cite{TZ} we introduced a model
for electrophoresis of polymers with
impurities, i.e., polymers having
segments (reptons) insensitive to an
external electric field $E$. We showed
that the effect of a single impurity is
equivalent to having a single kink
in the ASEP above. The drift velocity of
the polymer chain, being related in
this model to that of our random walker,
is therefore directly affected. The
details of this application, which is a
particular case of our model, are
presented in the last section of this
paper.

\vspace*{0.5cm}

The structure of the paper is as
follows: in Sect. 2 we provide a
precise formulation for the model
with one mobile directional impurity,
at the level of Master equations.
Here, we show that the structure of
these equations leads us to a natural
definition of a new entity, namely,
the quasi-walker,
which stands for the hole-DB couple.
A full solution for the steady state
probability distribution is then given.
From that, the expressions for the
velocity and diffusion constant
(in the steady state) of this quasi-walker
are derived. Section 3 is devoted
to a detailed description of the
original system in terms of the
quasi-walker. Using a technique
from the theory of random walks on rings
, see \cite{Derrida}, \cite{Hughes}
(and referred as replication method by us), the
expressions for the velocity and
diffusion constant for the hole and the
DB (with respect to the chain) are
obtained. The remaining sections deal
with special cases and applications.

\section{Definition of the model and the quasi-walker.}

In this section, we will present another 
representation of our model which
is equivalent to, but simpler than, that 
shown in Fig. 2. Though there are
two moving objects (the hole and the defect bond), 
the motion of the DB is
completely determined by the dynamics of 
the hole, so that there is,
effectively, only {\em one} degree of
freedom. This single co-ordinate is
then associated with the quasi-walker.

Instead of a chain on a kinked checkerboard, 
our system can be redrawn as in
Fig.3a)-d). Here, we have our fixed 
reference frame to be that of the
particle chain, arbitrarily choosing one 
particle as the ``first'' in the
labeling $1,2,...,N$. Representing the 
particles (hole) by solid (open)
circles, we see that the hole is located 
in between the particles, i.e., at
positions denoted by short vertical lines. 
The DB (which is the kink in
Fig.2) can also be thought of a ``wanderer'' 
amongst the particles. 
We denote its position by a long vertical line. 
The only complication arises
when the hole comes ``in contact'' with the 
DB, i.e., when both of them are
located between two successive particles. 
Then, there are two distinct
states of the system, depending on which 
side of the DB we find the hole. As
a result, we are forced to draw two short 
vertical lines on either side of the DB.

To keep track of these positions we 
introduce the following variables:\\ 
$\bullet$ Associate the position of the 
DB with $b$, i.e., when the defect
bond is between particles $b$ and $b+1$. 
Thus, $\ b\in [1,N]$. \\ 
$\bullet$ Let $m$ denote the position of 
the hole {\em relative to the} DB. Since the
hole occupies a square in the checkerboard, 
it is clear that this square can
be on either side of the DB (shown by 
solid line in Fig.2).  Thus, $m$
assumes one more value than $b$. We choose 
$\ m\in [0,N]$ and let $m=0$ and 
$N$ be associated with the positions 
next to the DB (See Fig.3). \\ 
For clarity, let us illustrate the dynamics 
of the hole near the DB. There are 4
representations of the same piece of chain 
at consecutive time steps. In the
first (Fig.3a), the hole is at $m=N-1$ and 
the bond is at $b$. In the next
step, the hole interchanges with particle 
$b$, arriving at $m=N$ (Fig.3b).
Suppose it next interchanges with 
particle $b+1$. In the checkerboard, this
would be an exchange ``across the DB'', 
so that the DB will now be found
between particles $b+1$ and $b+2$. At 
the same time, the hole will be ``on
the other side of the DB''. Thus, both 
the hole and the DB arrive at new
positions (Fig.3c). 
Finally, according to 
our scheme, we must ``relabel'' $m$
accordingly to the value $m=0$.
 In the last frame, 
the hole 
moves further to the right.
\begin{figure}
\vspace*{-1.5cm}
\hspace*{2.5cm} 
\epsfxsize = 4 in \epsfbox{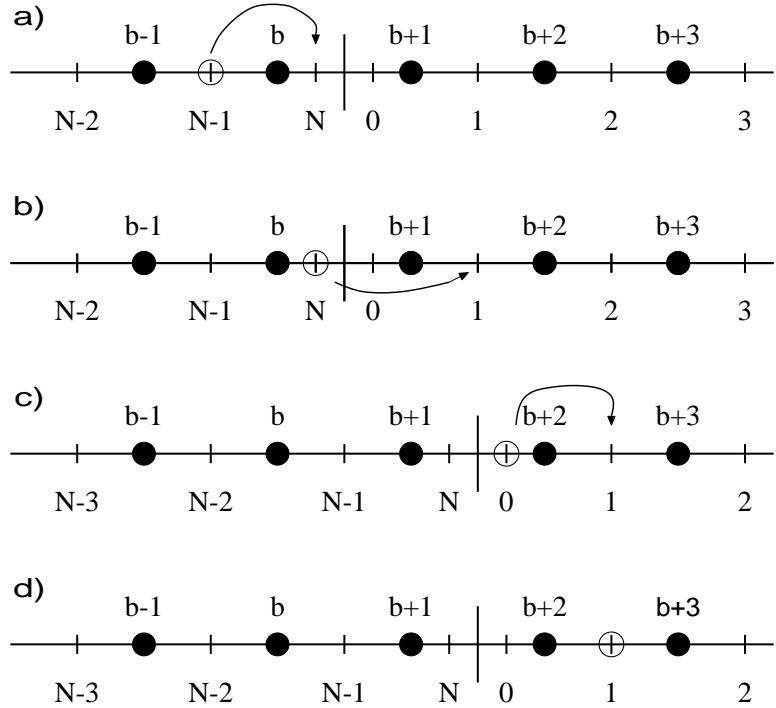}
\caption{An equivalent representation
of the model from Fig. 2. The solid circles 
represent particles. The short vertical lines
mark the possible positions of the hole represented
by an open circle. The DB is shown as a
long, solid, vertical line. As the hole
passes through the DB, the latter encounters
a one-step shift in the same direction, a)-d).}
\end{figure}
\noindent
The DB is not affected, so that only $m$ increases 
to $1$ (Fig.3d). Of course, if
the hole moves to the left, then we must 
reverse the previous process. Let
us emphasize that, in the single move between 
Figs.3b and 3c, $b$ increments
by $1$ while $m$ drops from $N$ to $0$. 
These rules can easily be verified
by playing on the checkerboard.

\subsection{The system of Master equations}

The full probabilistic information about 
our system is contained in $P(b,m;t)$, 
the probability to find the DB and the 
hole at $b$ and $m$, respectively,
at time $t$. The time evolution of $P$ 
will be controlled by the jump rates
of the hole, defined in rules 1)-5) above: 
$W_{i,j}$, $q$, and $q^{\prime }$. 
Recall that the subscripts on $W$ refer 
to the positions of the hole being
between particles $i$ and $i+1$, etc. 
In terms of our variables $b$ and $m$,
this position is clearly just $b+m$, modulo $N$, 
since our chain is a closed
one of $N$ particles. Thus, it is 
natural to introduce the notation: 
\begin{equation}
W_{m,m^{\prime }}^b\equiv 
W_{b+m,b+m^{\prime }}\;\;\;,  \label{newnot}
\end{equation}
with the understanding: 
\begin{equation}
W_{i+N,j+N}=W_{i,j} \quad .  \label{prdcty}
\end{equation}
Since the hole can only exchange with 
its neighboring particles, the
restriction $\mid \!m-m^{\prime }\!\mid =1$ 
applies in (\ref{newnot}).
Combining these remarks and following the 
special rules associated with the
DB (Fig.3), the evolution of $P$ is given 
in terms of a system of Master
equations:

\begin{eqnarray}
 {m=0}:\quad \partial _tP(b,0)
=\hspace*{-0.6cm}&&P(b-1,N)q+P(b,1)W_{0,1}^b 
\nonumber \\
&&-P(b,0)\Big(q^{\prime }+W_{1,0}^b\Big)  \nonumber \\
 {m\neq 0,N}:\quad \partial _tP(b,m)
=\hspace*{-0.6cm}&&P(b,m-1)W_{m,m-1}^b+
P(b,m+1)W_{m,m+1}^b  \nonumber \\
&&-P(b,m)\Big(W_{m-1,m}^b+W_{m+1,m}^b\Big)  \label{closed} \\
 {m=N}:\quad \partial _tP(b,N)
=\hspace*{-0.6cm}&&P(b,N-1)
W_{N,N-1}^b+P(b+1,0)q^{\prime }  \nonumber \\
&&-P(b,N)\Big(W_{N-1,N}^b+q\Big)  \nonumber
\end{eqnarray}\\
\noindent
where we have suppressed $t$ for simplicity.

\subsection{The steady state limit}

One of our central interests is the 
steady state limit. Assuming that it
exists, let us define\\ 
\begin{equation}
X_m^b\equiv \lim_{t\to \infty }
P(b,m;t)\;.  \label{steady}
\end{equation}\\
\noindent
Then, from (\ref{closed}), these 
satisfy the following set of linear
equations: \\
\begin{eqnarray}
{m=0}:\hspace*{-0.4cm}
&&X_1^bW_{0,1}^b+X_N^{b-1}q-X_0^b\Big( 
q^{\prime }+W_{1,0}^b\Big) =0  \nonumber \\
{m\neq 0,N}:\hspace*{-0.4cm}
&&X_{m+1}^bW_{m,m+1}^b+X_{m-1}^bW_{m,m-1}^b-
X_m^b \Big(W_{m-1,m}^b+W_{m+1,m}^b\Big)=0  \label{climit} \\
{m=N}:\hspace*{-0.4cm}
&&X_{N-1}^bW_{N,N-1}^b+X_0^{b+1}q^{\prime
}-X_N^b\Big(q+W_{N-1,N}^b\Big)=0  \nonumber
\end{eqnarray}\\
\noindent
These equations can be solved by 
introducing the steady state probability
current densities: \\
\begin{equation}
c_m^b\equiv X_m^bW_{m-1,m}^b-
X_{m-1}^bW_{m,m-1}^b\quad .  \label{cbm}
\end{equation}\\
\noindent
Now, one can rewrite (\ref{climit}) as: \\
\begin{eqnarray}
{m=0}:\qquad 
&&c_1^b=X_0^bq^{\prime }-X_N^{b-1}q\quad ,
\label{cclimit2} \\
{m\neq 0,N}:\qquad &&c_{m+1}^b=
c_m^b\quad ,\;\;\;\;\;  \label{cclimit1} \\
{m=N}:\qquad 
&&c_N^b=X_0^{b+1}q^{\prime }-X_N^bq\quad .
\label{cclimit3}
\end{eqnarray}\\
\noindent
From (\ref{cclimit1}), we find \\
\begin{equation}
c_N^b=c_{N-1}^b=...=c_2^b=c_1^b\quad .  \label{acb}
\end{equation}\\
\noindent
Examining the right hand sides 
of (\ref{cclimit2}) and (\ref{cclimit3}), we
can go one step further:\\ 
\begin{equation}
c_N^b=c_1^{b+1}\quad .  \label{acb1}
\end{equation}\\
\noindent
In other words, none of these $c$'s 
depend on $b$ or $m$, so that we might
as well write\\ 
\begin{equation}
c_m^b=c\quad .  \label{cbmeqc}
\end{equation}\\
\noindent
This equation can be easily interpreted 
as: the probability current density
in the steady state is a constant. 
This result is typical for systems with a
periodic, one dimensional configuration 
space, where the current has only a
single component and can be fixed by the 
conservation law. Since we have two
variables ($b,m$), we may regard our 
configuration space as ''two
dimensional'', so that (\ref{cbmeqc}) is 
perhaps surprising. The resolution
is best displayed by explicitly writing 
the entire system (\ref{cclimit2},
\ref{cclimit1}, \ref{cclimit3}) using 
(\ref{cbmeqc}), as in Fig.4. 
Note that, between the lowest and highest 
equations, we have the periodic
boundary condition (\ref{prdcty}), i.e., 
$X_N^0\equiv X_N^N$, and 
$X_0^{N+1}\equiv X_0^1$. 
Thus, the whole system of equations can be
regarded as those for 
\underline{a single, periodic chain}. Explicitly, we
introduce: \\
\begin{equation}
\overline{X}_k\equiv X_m^b,\;\;\;
b\in [1,N]\,,\;m\in [0,N]\quad ,
\label{rename}
\end{equation}\\
\noindent
where \\
\begin{equation}
k\equiv 1+(b-1)(N+1)+m  \label{renumber}
\end{equation}\\
\noindent
simply numbers the equations of Fig.4 
starting from the bottom. Thus, $k\in
[1,M]$, with \\
\begin{equation}
M=N(N+1)\quad .  \label{M}
\end{equation}\\
\noindent
Observe that $k$ {\em uniquely} determines 
the pair $(b,m)$ and vice-versa: \\
\begin{eqnarray}
b &=&\Big[ \frac{k-1}{N+1}\Big]+1\qquad \mbox{and} \\
& &\nonumber \\
m &=&(k-1)\;\;\mbox{mod}(N+1)\quad ,
\end{eqnarray}\\
\noindent
where $[x]$ represents the integer 
part of the real number $x$. To display
the structure of this single 
periodic chain more clearly, let us introduce:\\ 
\begin{equation}
\overline{W}_{k,k-1}\equiv W_{m,m-1}^b
\;\;\;\mbox{and}\;\;\;\overline{W}_{k-1,k}
\equiv W_{m-1,m}^b\quad ,
\end{equation}\\
\noindent
for $m\in [1,N].$ For $m=0$, we first define\\ 
\begin{equation}
k_b\equiv 1+(b-1)(N+1)\quad .  \label{kabe}
\end{equation}\\
\noindent
Referring to (\ref{cclimit2}, 
\ref{cclimit3}), we see that \\
\begin{equation}
\overline{W}_{k_b,k_b-1}=q\;\;\;
\mbox{and}\;\;\;\overline{W}_{k_b-1,k_b}=q^{\prime }\quad ,
\end{equation}\\
\noindent
for all $b$. Finally, to stress the 
$M$-periodicity of these rates, we write \\
\begin{equation}
\overline{W}_{k+M,l+M}=\overline{W}_{k,l}\quad .  \label{22}
\end{equation}\\
\noindent
With this notation the system from Fig.4 reads:\\ 
\begin{equation}
\overline{X}_k\overline{W}_{k-1,k}-\overline{X}_{k-1}\overline{W} 
_{k,k-1}=c\;,\;\;\;k\in [1,M]\quad .  \label{25}
\end{equation}

\begin{figure}
\vspace*{-2cm}
\hspace*{-1cm}\epsfxsize = 8 in \epsfbox{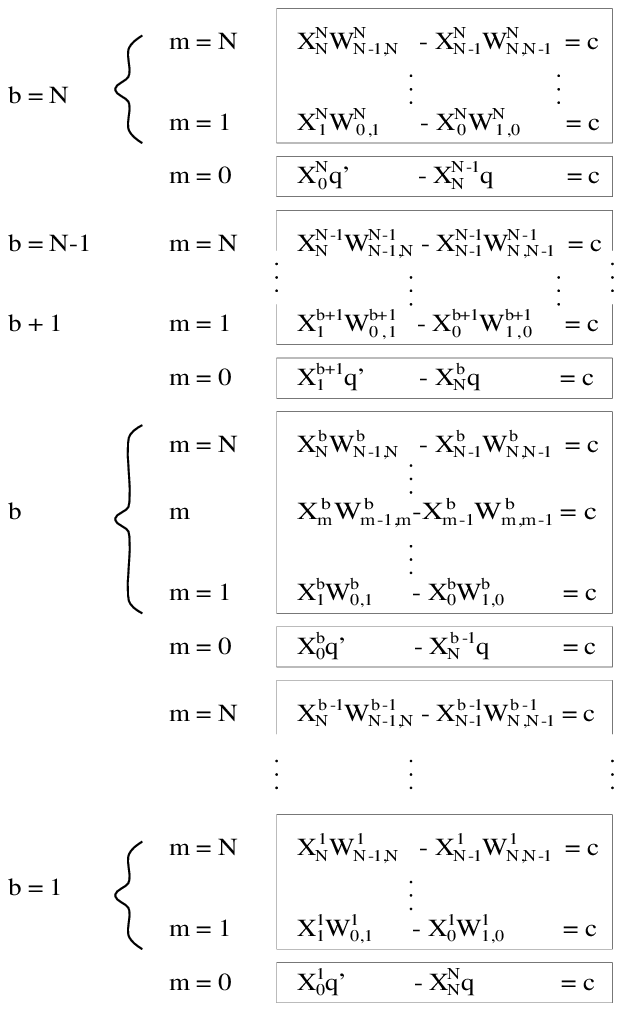}
\caption{ The rewritten system. $b$ labels
the blocks, $m$ the corresponding equations.}
\end{figure}
 
\noindent
Recast in the form of (\ref{25}), our 
system is exactly the one studied by
Derrida, \cite{Derrida}. The only 
difference is that, in our case, the 
$\overline{X}_k$ stands for the steady 
state of the ``defect bond-hole''
pair, as opposed to a single particle 
as in \cite{Derrida}. We will refer to
this new entity as a ``quasi-walker'', 
and treat it as a single random
walker on a chain of length $M$.

Lastly, we remark that not only the steady 
state limit can be reduced to the
case in \cite{Derrida}, but also the 
full time dependence of our system (\ref
{closed}) can be mapped. Defining \\
\begin{equation}
\overline{P}_k(t)\equiv P(b,m;t)\quad ,\;\;  \label{relab}
\end{equation}\\
\noindent
the system of Master equations 
(\ref{closed}) can be written as: \\
\begin{equation}
\partial _t\overline{P}_k=\overline{P}_{k+1}
\overline{W}_{k,k+1}+\overline{P}_{k-1}
\overline{W}_{k,k-1}-\overline{P}_k\Big(\overline{W}_{k+1,k}+
\overline{W}_{k-1,k}\Big)\quad .  \label{26} \\
\end{equation}

\subsection{Application of Derrida's 
results to the quasi-walker.}

In this paragraph we apply directly 
the results from \cite{Derrida} to
obtain the steady state probability 
distribution, the velocity and diffusion
constants of our quasi-walker. 
As expected, all quantities are rational
functions of the jump rates. Thus,\\ 
\begin{equation}
\overline{X}_k=\Gamma \overline{r}_k,  \label{27}
\end{equation}\\
\noindent
where \\
\begin{equation}
\overline{r}_k\equiv \frac 1{\overline{W}_{k+1,k}}\left[
1+\sum_{i=1}^{M-1}\prod_{j=1}^i
\frac{\overline{W}_{k+j-1,k+j}}
{\overline{W}_{k+j+1,k+j}}\right] \quad ,  \label{28}
\end{equation}\\
\noindent
and $\Gamma $ is a normalization constant: \\
\begin{equation}
\Gamma =\Big( \sum_{k=1}^M
\overline{r}_k\Big)^{-1}.  \label{29}
\end{equation}\\
\noindent
The velocity and the diffusion constant are given, respectively, by: \\
\begin{equation}
V=\Gamma M\left[ 1-\prod_{k=1}^M\frac{\overline{W}_{k,k+1}}
{\overline{W}_{k+1,k}}\right] \quad ,  \label{34}
\end{equation}\\
\noindent
and \\
\begin{equation}
D=\Gamma ^2\left[ V\sum_{k=1}^M
\overline{u}_k\sum_{i=1}^Mi\overline{r}_{k+i}+M
\sum_{k=1}^M\overline{W}_{k+1,k}
\overline{u}_k\overline{r}_k\right]
-V\frac{M+2}2\quad ,  \label{diffc}
\end{equation}\\
\noindent
where \\
\begin{equation}
\overline{u}_k\equiv \frac 1
{\overline{W}_{k+1,k}}\left[
1+\sum_{i=1}^{M-1}\prod_{j=1}^i
\frac{\overline{W}_{k-j,k+1-j}}{
\overline{W}_{k+1-j,k-j}}\right] \quad .  \label{uk}
\end{equation}\\

While it is straightforward to arrive 
at (\ref{25}-\ref{29}), a subtle
process is involved in finding 
(\ref{34},\ref{diffc}). Before proceeding,
let us review this method 
\cite{Derrida}, which we call ''replication''.
Instead of studying a finite chain 
with periodic boundary conditions,
replicate it into an infinite, 
periodic chain. Thus, we allow, for example,
the indices $i,j$ in $\overline{W}_{i,j}$ 
to be {\em any} integer, together
with the periodicity constraint, 
(\ref{22}). Similarly, the distribution 
$\overline{P}_k(t)$ in (\ref{relab}) 
is now defined for all $k\in [-\infty,
\infty ]$, and satisfies (\ref{26}) 
with (\ref{22}). To compute the steady
state quantities defined for the 
original chain, we must sum over all
equivalent sites on the replicated 
chain. On the other hand, the velocity
and diffusion constants can be found 
from the first and second order moments
of $\overline{P}_k$. Introducing \\
\begin{equation}
 x(t)  =\sum_{k=-\infty }^\infty 
k\overline{P}_k(t)\quad \mbox{and}\quad
 x_2(t)  =\sum_{k=-\infty }^\infty 
k^2\overline{P}_k(t)\;\quad ,  \label{mom}
\end{equation}\\
\noindent
the velocity is naturally \\
\begin{equation}
V=\lim_{t\to \infty }\frac d{dt} 
x(t)  \quad ,  \label{30}
\end{equation}\\
\noindent
while the diffusion constant is given by: \\
\begin{equation}
D=\lim_{t\to \infty }\frac d{dt}
\big[  x_2(t) - 
 (x(t))^2 \big]\quad .
\label{31}
\end{equation}\\
\noindent
Note that $x(t)$ is the position 
of the quasi-walker at time $t$, averaged
over all paths. In Sect. 3 we present 
in more detail an extension of this
replication technique to describe the 
hole and the defect bond separately.

Before proceeding, let us present 
these results in terms of the original
jump rates $W_{i,j}$, $q^{\prime }$ 
and $q$. For convenience, let us first
define the following expressions: \\
\begin{equation}
S(b,m)\equiv \sum_{i=1}^{N-m-1}
\prod_{l=1}^i\frac{W_{b+m+l-1,b+m+l}}
{W_{b+m+l+1,b+m+l}}\;,
\qquad m\in [0,N-2]\quad ,  \label{svb}
\end{equation}\\
\noindent
\begin{equation}
S(b,N-1)\equiv 0\quad ,  \label{svbn}
\end{equation}\\
\noindent
\begin{equation}
T(b,m)\equiv \sum_{i=1}^m
\prod_{l=1}^i\frac{W_{b+m-l,b+m+1-l}}
{W_{b+m+1-l,b+m-l}}\;,\qquad m\in [1,N]\quad ,  \label{tvb}
\end{equation}\\
\noindent
\begin{equation}
T(b,0)\equiv 0\quad .  \label{tvb0}
\end{equation}\\
\noindent
The last term of (\ref{svb}) and 
(\ref{tvb}) also appear frequently, so that
we define: \\
\begin{equation}
\lambda _S(b,m)\equiv \prod_{l=1}^
{N-m-1}\frac{W_{b+m+l-1,b+m+l}}
{W_{b+m+l+1,b+m+l}}\quad ,\qquad m\in [0,N-2]\quad ,
\end{equation}\\
\noindent
\begin{equation}
\lambda _S(b,N-1)\equiv 0\quad ,
\end{equation}\\
\noindent
\begin{equation}
\lambda _T(b,m)\equiv \prod_{l=1}^m
\frac{W_{b+m-l,b+m-l+1}}{W_{b+m-l+1,b+m-l}}
\quad ,\qquad m\in [1,N]\quad ,
\end{equation}\\
\noindent
\begin{equation}
\lambda _T(b,0)\equiv 1\quad .
\end{equation}\\
\noindent
Observe that the above quantities 
depend on neither $q^{\prime }$ nor $q$,
and that the $W$'s satisfy periodicity 
(\ref{prdcty}). As a result, they are
also periodic: e.g., $S(b+N,m)=S(b,m),$ etc.

Finally, let: \\
\begin{equation}
\lambda \equiv \prod_{b=1}^N
\frac{W_{b-1,b}}{W_{b+1,b}}\quad .  \label{l}
\end{equation}\\
\noindent
and \\
\begin{equation}
U\equiv 1-\Big( \frac{q^{\prime }}q
\lambda \Big)^N\quad .  \label{v}
\end{equation}\\

After some rather tedious algebra, we 
arrive at the explicit expressions:\\ 
\begin{equation}
\overline{r}_{k_b+m}=\frac{1}
{W_{b+m+1,b+m}}\left\{ U\big[1+S(b,m)\big]
+W_{b-1,b}\lambda _S(b,m)\overline{r}_{k_b+N}\right\} ,  \label{70}
\end{equation}\\
\noindent\\
\begin{eqnarray}
\overline{u}_{k_b+m}=\frac{1}
{W_{b+m+1,b+m}}\left\{ U\big[1+T(b,m)\big]
+q^{\prime }\lambda _T(b,m)
\overline{u}_{k_{b-1}+N}\right\} && , \\
m\in [0,N-1]&&  \nonumber  \label{hetvenegy}
\end{eqnarray}\\
\noindent
where \\
\begin{equation}
\overline{r}_{k_b+N}=\frac{1}q
\sum_{i=1}^N\Big( \frac{q^{\prime }}q\lambda 
\Big)^{i-1}\left\{ 1+q^{\prime }\frac 1{W_{b+i+1,b+i}}\left[
1+S(b+i,0)\right] \right\}   \label{60}
\end{equation}\\
\noindent
and \\
\begin{equation}
\overline{u}_{k_b+N}=\frac 1q\sum_{i=0}^{N-1}
\Big( \frac{q^{\prime }}q
\lambda \Big)^i\big[1+T(b-i,N)\big]\quad .  \label{62}
\end{equation}\\
\noindent
The normalization constant (\ref{29}) now takes the form: \\
\begin{equation}
\Gamma^{-1} =\sum_{b=1}^N\left\{ U\ 
\sum_{m=0}^{N-1}\frac{1+S(b,m)}{W_{b+m+1,b+m}
}+\left[ 1+W_{b-1,b}\sum_{m=0}^{N-1}
\frac{\lambda _S(b,m)}{W_{b+m+1,b+m}}
\right] \overline{r}_{k_b+N}.\right\}\quad .  \label{91}
\end{equation}\\
\noindent
From (\ref{25}), we see that the current 
$c$ can be related \cite{Derrida} to
the velocity of the quasi-walker via : \\
\begin{equation}
c=-\frac VM\quad ,  \label{92}
\end{equation}\\
\noindent
with $V$ given by (\ref{34}). Verifying\\ 
\begin{equation}
\prod_{k=1}^M\frac{\overline{W}_{k,k+1}}
{\overline{W}_{k+1,k}}=\left( 
\frac{q^{\prime }}q\lambda \right) ^N\quad ,  \label{93}
\end{equation}\\
\noindent
we have: \\
\begin{equation}
V=\Gamma MU\quad .  \label{94}
\end{equation}\\
\noindent
In other words, $U$ from (\ref{v}) contains 
the essential information on the
velocity $V$. Further, it indicates whether 
our system satisfies detailed
balance or not, i.e., if we have an 
equilibrium or a non-equilibrium steady
state. The former is the case if and 
only if $c=0$, so that both $U$ and $V$
vanishes. The condition to {\em stop} 
the motion can be summarized in a
simple equation: \\
\begin{equation}
q/q^{\prime }=\lambda \quad .  \label{95}
\end{equation}\\
\noindent
Its interpretation is clear. The left hand 
side is the bias in the jump rate
across the defect bond, while the right 
hand side represents the bias in the
opposite direction across the {\em entire chain } 
of particles. When these
balance, we have equilibrium.

In the next section, we will see 
that the expression (\ref{94}) for the
velocity of the quasi-walker plays 
a central role in determining the
velocities of the DB ($V_{bond}$) 
and the hole ($V_{hole}$) with respect to
the chain, as well as the relative 
velocity between the two ($V_{rel}$). In
particular, an extension of the 
method of replicating a finite closed chain
will be shown in the derivation of 
these quantities. Following this method
we also can derive the diffusion 
constants for the bond and the hole 
$D_{bond}$ and $D_{hole}$ in the same steady state limit.

\section{Replicating the finite, closed chain}

In a finite, closed chain, a hole 
will visit the same site infinitely many
times in the $t\rightarrow \infty$ 
limit. As a result, it is not easy to
find asymptotic properties like steady 
state velocities and diffusion
constants. To circumvent these 
difficulties, the method used (see for e.g.
\cite{Derrida}, \cite{Hughes}) is 
to replicate the closed
chain into an infinite, periodic chain
(with periodic jumps rates).
Since there are infinitely many
sites, though many are ``equivalent'',
this infinite chain greatly
facilitates the computation of
asymptotic quantities. Here, we will
generalize this method and arrive at 
expressions for the velocities and
diffusion constants of both the DB and 
the hole. The extra complication in
our case is that, instead of having 
one random walker, we have two moving
objects: the hole and the defect bond. 
Therefore, we have to define our
replicated model carefully so as not 
to lose any physical information about
the finite, closed chain.

On the infinite chain, the position
of both the hole and the DB can take
arbitrary values, so that they can be 
found with any distance in between.
Let us define the position of the DB, 
relative to some arbitrarily chosen
``origin'', to be $\beta $ and the
distance from the bond to the hole to be 
$\mu $. Analogous to $b$ and $m$, 
these differ in that $\beta ,\mu \in 
\left[ -\infty ,\infty \right] \quad$ .  

Now, the dynamical rules associated 
with these are not so simple, however,
since they must reflect the fact that, 
on the closed finite chain, the DB
``advances'' by a step {\em each time} 
the hole moves forward $N$ steps. To
simulate these advances, we introduce 
an infinite set of ``images'' of the
bond, located at multiples of $N$ from
the DB. These images serve as
``condition lines'' for the motion of 
the DB itself, i.e., whenever the hole
passes through {\em one} of these images, 
the $\beta$ changes by unity. Of
course, in accordance with the idea of 
an image, {\em all} the images also
move correspondingly. In this way, 
though the two objects can be arbitrarily
far apart on the replicated chain, the 
properties, or the structure, of the
original chain are replicated faithfully.
Referring to Fig.5, it is clear that 
there are unique integers $i$ and $j$,
such that\\
\begin{equation}
\beta =b+iN \qquad \mbox{and\qquad }
\mu =m+j(N+1)\quad .  \label{bebmum}
\end{equation}\\
\begin{figure}
\vspace*{-3.5cm}
\epsfxsize = 6 in \epsfbox{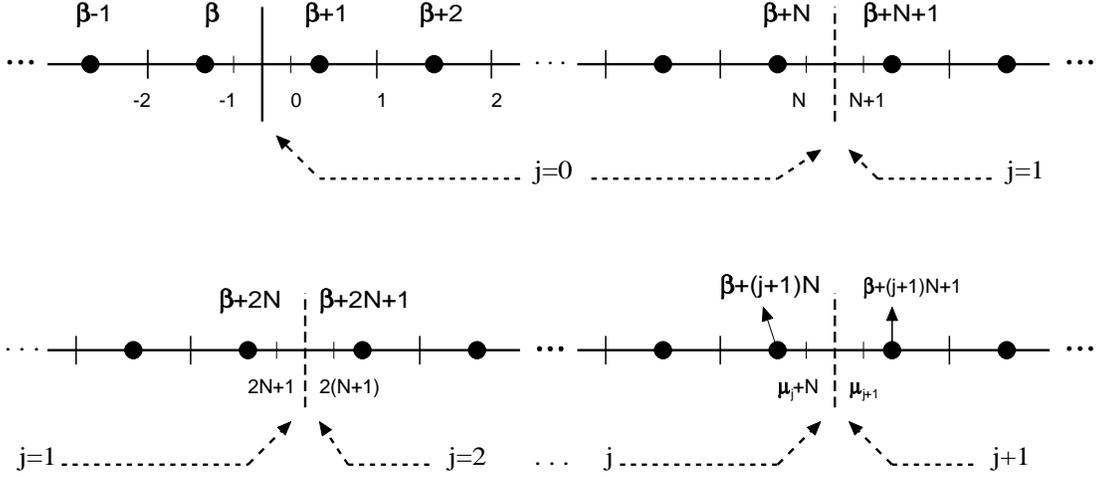}
\vspace*{-3.5cm}
\caption{The replicated, infinite chain.
The long, solid vertical line represents
the DB's position while the long, dashed
vertical ones show its images. $j$ labels
the number of complete blocks (containing
$N$ particles) away from the DB.}
\end{figure}
\noindent
Note that, as expected, $b$ is
just $\beta\;\;mod\;\;(N)$. Now, $j$ represents
the number of {\em complete} replicas
of the original chain lying between
the DB and the hole, while $m$ is
just the distance, as in the previous
section, from the hole to the nearest
{\em bond-image} to its left. Let us
caution the reader on an extra
complication associated with the second
equation in (\ref{bebmum}). The reason
for the factor $(N+1)$ lies in our
having to distinguish between a hole
immediately to the left/right of all
the images of the DB. Therefore $\mu$
represents the {\em number of states
or positions} between the bond and the hole.
However, the ``actual distance'', in
terms of {\em number of particles}
(or units of lattice constant),
between the DB and the hole is
only $m+jN=\mu -j$ (see Fig.5).
In other words, while
it is necessary to associate
two positions with each image for the hole,
there is only one ``lattice spacing''. This subtle
distinction will be crucial when we
compute the velocity and diffusion
constants, both of which must be
based on the physical distance. In
particular, the distance of the hole
from the origin of the infinite chain
is simply $\beta +\mu -j.$

To distinguish the statistical
mechanics of the replicated chain from the
finite one, we will use notation
$\tilde{P}(\beta ,\mu ;t)$ for the
probability distribution. To write
down the Master equations for $\tilde{P}$
, based on the original set (\ref{closed}),
we first transcribe $\tilde{W}
_{\mu ,\mu ^{\prime }}^\beta $,
the transition probability for hole-jumps
on the replicated chain. Using the
periodicity property (\ref{prdcty}), we
see that these are simply given by\\
\begin{equation}
\tilde{W}_{\mu ,\mu ^{\prime }}^\beta
=W_{m,m^{\prime }}^b\quad . \label{pula}
\end{equation}\\
\noindent
With the notation $\mu _j\equiv \ j(N+1)$,
and suppressing $t$, we define the currents\\
\begin{eqnarray}
\qquad  {m\neq 0}:\quad
\tilde{{\cal J}}_\mu^{\beta} (\tilde{P})
\equiv\hspace*{-0.6cm}&&\tilde{P}(\beta ,\mu )
\tilde{W}_{\mu-1,\mu}^{\beta}-\tilde{P}
(\beta ,\mu-1)\tilde{W}_{\mu ,
\mu-1}^\beta,  \nonumber \\
\qquad  {m=0}:\qquad
\tilde{{\cal J}}_{\mu _j}^{\beta} (\tilde{P})
\equiv \hspace*{-0.6cm}&&\tilde{P}(\beta ,\mu{_j})
q^{\prime }-\tilde{P}(\beta-1,\mu
_{j-1}+N)q\quad .  \label{ketto}
\end{eqnarray}\\
\noindent
In terms of these, the system of
Master equations for $\tilde{P}
(\beta ,\mu)$ reads as:\\
\begin{eqnarray} \mbox{
\parbox{10cm}{\begin{eqnarray}
 {m=0}:  \qquad \qquad
\partial_t \tilde{P}(\beta,\mu_j)=\hspace*{-0.6cm}&&
\widetilde{{\cal J}}^{\beta}_{\mu_j+1}(\tilde{P})
-\widetilde{{\cal J}}^{\beta}_
{\mu_j}(\tilde{P}) \nonumber \\
 {m\neq 0,N}: \;\;\;\; \qquad
\partial_t \tilde{P}(\beta,\mu)=\hspace*{-0.6cm}&&
\widetilde{{\cal J}}^{\beta}_{\mu+1}(\tilde{P})
-\widetilde{{\cal J}}^
{\beta}_{\mu}(\tilde{P}) \nonumber \\
\qquad \quad  {m=N}:  \qquad
\partial_t \tilde{P}(\beta,\mu_j+N)=\hspace*{-0.6cm}&&
\widetilde{{\cal J}}^{\beta+1}_{\mu_{j+1}}(\tilde{P})
-\widetilde{{\cal J}}^
{\beta}_{\mu_j+N}(\tilde{P}). \nonumber
\end{eqnarray}} } \label{harom} \end{eqnarray}

\subsection{Positions, velocities
and diffusion constants}

To reach our goal of computing
velocities, etc., we need some notion of
positions. For that, we introduce
the following first moments of $\tilde{P}$:\\
\begin{equation}
\langle \beta(t) \rangle=\sum_{\mu =-
\infty}^{\infty}\sum_{\beta=-\infty
}^{\infty}\beta \tilde{P}(\beta,\mu)\quad ,  \label{ot}
\end{equation}\\
\noindent\\
\begin{equation}
\langle \mu(t) \rangle =\sum_{\mu =-\infty }^
{\infty} \sum_{\beta =-\infty }^{\infty}
\mu \tilde{P}(\beta,\mu) \quad ,  \label{hat}
\end{equation}\\
\noindent\\
\begin{equation}
\langle j(t) \rangle =\sum_{\mu =-\infty }^{\infty}
 \sum_{\beta =-\infty }^{\infty} j
\tilde{P}(\beta,\mu)\quad .  \label{hatvesz}
\end{equation}\\

Is easy to see that, the average position
for the DB is just $\langle \beta(t) 
\rangle.$ For the average distance
between the hole and the DB, we define\\
\begin{equation}
\langle \delta(t) \rangle=\langle \mu(t) \rangle
-\langle j(t) \rangle\quad ,  \label{tav}
\end{equation}\\
\noindent
so that the average position of the hole is\\
\begin{equation}
\langle h(t) \rangle \equiv \langle \beta(t) \rangle
+\langle \delta(t) \rangle=\langle \beta(t) \rangle
+\langle \mu(t) \rangle-
\langle j(t) \rangle\quad .  \label{hate}
\end{equation}\\

The time derivatives of the moments
defined above can be used to compute
velocities:\\
\begin{eqnarray}
v_{bond}(t) &\equiv &\frac{d}{dt}
\langle \beta(t) \rangle \quad ,  \label{vebo} \\
v_\mu(t) &\equiv &\frac{d}{dt}
\langle \mu(t) \rangle\quad ,  \label{vequ} \\
v_j(t) &\equiv &\frac{d}{dt}
\langle j(t) \rangle\quad . \label{vete}
\end{eqnarray}\\
\noindent
The first clearly represents the
average velocity, with respect to the
chain, of the DB. Naively, the second
can be thought of as the velocity of
the hole relative to the bond on this
infinite chain, since $\mu$ plays the
role of $m$ here. However, in view of
the distinction between $\mu $ and the
``actual distance'' $\mu -j$, we
emphasize that the ``actual velocity'' of
the hole relative to the bond, is\\
\begin{equation}
v_{rel}(t)\equiv \frac{d}{dt}\langle 
\delta(t) \rangle=v_\mu (t)-v_j(t)\quad ,
\label{verela}
\end{equation}\\
\noindent
while the hole's velocity, relative to
the chain itself is\\
\begin{equation}
v_{hole}\equiv \frac{d}{dt}\langle h(t) \rangle
=v_{bond}(t)+v_{rel}(t)\quad .
\label{veho}
\end{equation}\\

The diffusion constants for the DB and
for the hole can be found similarly:\\
\begin{equation}
d_{bond}(t)=\frac{1}{2}[\frac{d}{dt}
(\langle \beta^{2}(t) \rangle)-\frac{d}{dt}(
\langle \beta(t) \rangle^{2})]\quad ,  \label{kilenc}
\end{equation}\\
\noindent\\
\begin{equation}
d_{hole}(t)=\frac{1}{2}[\frac{d}{dt}
(\langle h^{2}(t) \rangle )-\frac{d}{dt}
(\langle h(t) \rangle^2)]\quad ,  \label{kilencvesz}
\end{equation}\\
\noindent
where\\
\begin{equation}
\frac{d}{dt}[\langle \beta^{2}(t) \rangle ]=
\sum_{\mu =-\infty }^{\infty}\sum_{\beta
=-\infty }^{\infty}\beta^{2}\partial _t
\tilde{P}(\beta ,\mu )\quad ,
\label{tizenegy}
\end{equation}\\
\noindent\\
\begin{equation}
\frac{d}{dt}[\langle h^{2}(t) \rangle]=\sum_
{\mu =-\infty }^{\infty} \sum_{\beta
=-\infty}^{\infty}(\beta+\mu-j)^{2}
\partial _t\tilde{P}(\beta,\mu)\quad .
\label{tizenketto}
\end{equation}\\
\noindent
Needless to say, all the six second
order moments are needed (all quadratic
combinations of $\beta$, $\mu$ and $j$).
In this paper, we will show
explicit calculations for the first
order moments and velocities only, since
the computations for the second order
moments are considerably lengthier yet
providing no additional insight.

\subsection{Reductions to the finite, closed chain}

As the original system consists
of only a finite, closed chain, it is
important to make the connection
between the replicated, infinite chain
above and the finite one. The simplest
way is to note that $i$ and $j$ play
the role of replica indices. Thus, we have:\\
\begin{equation}
P(b,m;t)=\sum_{i=-\infty }^{\infty}
\sum_{j=-\infty }^{\infty}\tilde{P}(\beta
,\mu ;t)\quad .  \label{tizenharom}
\end{equation}\\
\noindent
To facilitate the discussion of
the moments (\ref{ot}-\ref{hatvesz}), we
define\\
\begin{equation}
\phi (b,m)=\sum_{i=-\infty }^{\infty}
\sum_{j=-\infty }^{\infty} \beta \tilde{P}
(\beta ,\mu )\quad ,  \label{tizennegy}
\end{equation}\\
\noindent\\
\begin{equation}
\psi (b,m)=\sum_{i=-\infty }^{\infty}
\sum_{j=-\infty }^{\infty} \mu \tilde{P}
(\beta ,\mu )\quad ,  \label{tizenot}
\end{equation}\\
\noindent
so that the positions and
velocities are given by\\
\begin{equation}
\langle \beta(t) \rangle=\sum_{b=1}^N
\sum_{m=0}^N\phi(b,m)\quad ,
\label{tizenhat}
\end{equation}\\
\noindent\\
\begin{equation}
\langle \mu(t) \rangle=\sum_{b=1}^N
\sum_{m=0}^N\psi(b,m)\quad ,  \label{tizhet}
\end{equation}\\
\noindent\\
\begin{equation}
\langle j(t) \rangle=\frac{1}{N+1}\sum_{b=1}^N\sum_{m=0}^N
\left[\psi (b,m)-mP(b,m)\right] =
\frac{1}{N+1}\left[ \langle \mu(t) \rangle-\sum_{b=1}^N
\sum_{m=0}^{N}mP(b,m) \right] \quad ,  \label{tizhot}
\end{equation}\\
\noindent\\
\begin{equation}
v_{bond}(t)=\sum_{b=1}^N\sum_{m=0}^N
\partial _t\phi (b,m)\quad ,
\label{tinhut}
\end{equation}\\
\noindent\\
\begin{equation}
v_\mu (t)=\sum_{b=1}^N\sum_{m=0}^N
\partial _t\psi (b,m)\quad ,
\label{tinhit}
\end{equation}\\
\noindent
and\\
\begin{equation}
v_j(t)=\frac{1}{N+1}\left[ v_\mu(t)-
\sum_{b=1}^N \sum_{m=0}^Nm \partial
_tP(b,m) \right] .  \label{tunhit}
\end{equation}\\
\noindent
To compute (\ref{tinhut}-\ref{tunhit})
and the time derivatives of the six
second order moments, we should use
the Master equation for $P(b,m)$ to
eliminate $\partial _tP$ in favor of
the currents and derive similar
equations for $\phi (b,m)$ and
$\psi(b,m)$. Not surprisingly, these
equations will comprise of current-like
terms. Letting $\alpha$ be any of
the quantities $P$, $\phi$ or $\psi$, we define:\\
\begin{eqnarray}
\mbox{
\parbox{10cm}{\begin{eqnarray}
\qquad \qquad  {m=0}:
\;\;\;{\cal J}^{b}_{0} (\alpha) \equiv\hspace*{-0.6cm}&&
\alpha(b,0)q'-
\alpha(b-1,N)q \nonumber \\
\qquad \qquad  {m \neq 0}: \;\;\;
{\cal J}^
{b}_{m} (\alpha) \equiv\hspace*{-0.6cm}&&
\alpha(b,m)W^{b}_{m-1,m}-
\alpha(b,m-1)W^{b}_{m,m-1}, \nonumber
\end{eqnarray}}  }  \label{huszonharom}
\end{eqnarray}\\
\noindent
and note that these represent the
{\em net} probability current density from
state $(b,0)$ to $(b-1,N)$ for $m=0$
and $(b,m)$ to $(b,m-1)$ for $m\neq 0$,
respectively. Using these, the
definitions (\ref{tizenharom}-\ref{tizenot})
and the Master equation for
$\tilde{P}$, we obtain:
\begin{eqnarray}
\mbox{
\parbox{11cm}{\begin{eqnarray}
\noindent
 { m = 0:} \qquad \qquad
\partial_t  P(b,0) =\hspace*{-0.6cm}&& 
{\cal J}^{b}_{1}(P)-
{\cal J}^{b}_{0}(P) \nonumber \\
 {0< m < N:} \qquad \qquad
\partial_t  P(b,m)=\hspace*{-0.6cm}&& 
{\cal J}^{b}_{m+1}(P)-
{\cal J}^{b}_{m}(P) \nonumber \\
 { m = N:} \qquad \qquad
\partial_t  P(b,N)=\hspace*{-0.6cm}&& 
{\cal J}^{b+1}_{0}(P)
-{\cal J}^{b}_{N}(P) \nonumber
\end{eqnarray}} }\quad ,  \label{huszonnegy}
\end{eqnarray}
\begin{eqnarray}
\mbox{
\parbox{11cm}{\begin{eqnarray}
\noindent
 { m = 0:} \qquad \qquad
\partial_t  \phi(b,0)=\hspace*{-0.6cm}&& 
{\cal J}^{b}_{1}(\phi)-
{\cal J}^{b}_{0}(\phi) +P(b-1,N)q \nonumber \\
{0< m < N:} \qquad \qquad
\partial_t  \phi(b,m)=\hspace*{-0.6cm}&& 
{\cal J}^{b}_{m+1}(\phi)-
{\cal J}^{b}_{m}(\phi) \nonumber \\
{ m = N:} \qquad \qquad
\partial_t  \phi(b,N)=\hspace*{-0.6cm}&& 
{\cal J}^{b+1}_{0}(\phi)
-{\cal J}^{b}_{N}(\phi)-P(b+1,0)q' \nonumber
\end{eqnarray}} }\quad ,  \label{huszonot}
\end{eqnarray}
\begin{eqnarray}
\mbox{
\parbox{11cm}{\begin{eqnarray}
\noindent
{ m = 0:} \qquad
\partial_t  \psi(b,0)=\hspace*{-0.6cm}&& 
{\cal J}^{b}_{1}(\psi)-
{\cal J}^{b}_{0}(\psi)
+P(b-1,N)q-P(b,1)W^b_{0,1} \nonumber \\
{0< m < N:} \qquad
\partial_t  \psi(b,m)=\hspace*{-0.6cm}&& 
{\cal J}^{b}_{m+1}(\psi)-
{\cal J}^{b}_{m}(\psi)
+ P(b,m-1)W^{b}_{m,m-1}- \nonumber \\
&& P(b,m+1)W^{b}_{m,m+1} \nonumber \\
 { m = N:} \qquad
\partial_t  \psi(b,N)=\hspace*{-0.6cm}&& 
{\cal J}^{b+1}_{0}(\psi)
-{\cal J}^{b}_{N}(\psi)
+ P(b,N-1)W^{b}_{N,N-1}- \nonumber \\
& &P(b+1,0)q' \nonumber
\end{eqnarray}}  }\qquad .  \label{huszonhat}
\end{eqnarray}
Note that (\ref{huszonnegy}) is just
(\ref{closed}), as it should be. Now,
keeping in mind the periodicity property
$P(b+N,m)=P(b,m)$, we may express
for the velocities (\ref{tinhut}-
\ref{tunhit}) in terms of the probability
current-density alone:\\
\begin{eqnarray}
v_{bond}(t)=\hspace*{-0.6cm}&&-\sum_{b=1}^N
{\cal J}_0^b(P)\quad ,  \label{huszonhet} \\
\nonumber \\
v_\mu (t)=\hspace*{-0.6cm}&&-\sum_{b=1}^N\sum_{m=0}^N
{\cal J}_m^b(P)\quad ,\label{huszonnyolc} \\
\nonumber \\
v_j(t)=\hspace*{-0.6cm}&&v_{bond}(t)\quad.  \label{vbond=vj}
\end{eqnarray}\\
\noindent
The last of these, obtained by
explicit computation, can be understood
intuitively. Since $j$ measures of the
{\em net} number of times the hole
traverse around the (finite) chain,
while the DB moves one step each time
the hole crosses it, the displacements
in $j$ and the bond are locked.
From these, we easily arrive at\\
\begin{equation}
v_{rel}(t)=v_\mu (t)-v_j(t)=
v_\mu(t)-v_{bond}(t)  \label{hun??}
\end{equation}\\
\noindent
and\\
\begin{equation}
v_{hole}(t)=v_\mu (t).  \label{hunyi}
\end{equation}
\\
The $t$-derivative of the second
moments, (\ref{tizenegy}) and (\ref
{tizenketto}) can be computed similarly:\\
\begin{equation}
\frac{d}{dt}(\langle \beta^2(t) \rangle)=
-2\sum_{b=1}^N{\cal J}_0^b(\phi
)+\sum_{b=1}^N[P(b,N)q+P(b,0)
q^{\prime }].  \label{harmincketto}
\end{equation}\\
\noindent
and\\
\begin{eqnarray}
&&\frac{d}{dt}(\langle h^2(t) \rangle)
=-2\sum_{b=1}^N\sum_{m=0}^N{\cal J}_m^b(\phi
)-2\frac N{N+1}\sum_{b=1}^N
\sum_{m=0}^N{\cal J}_m^b(\psi )  \nonumber \\
&&\qquad \qquad \qquad -\frac{2}
{N+1}\sum_{b=1}^N\sum_{m=0}^Nm{\cal J}
_m^b(P)+2\sum_{b=1}^NP(b-1,N)q  \nonumber \\
+\sum_{b=1}^N\hspace*{-0.6cm}&&\Big\{ [P(b,0)q^{\prime
}+P(b-1,N)q]+\sum_{m=1}^N[P(b,m)W_{m-1,m}^b
+P(b,m-1)W_{m,m-1}^b]\Big\} .\label{harmincharom}
\end{eqnarray}\\
\noindent
Note that these expressions involve
sums, as opposed to differences, of
terms like $Pq$ and $PW$, so that
they are not simply related to the currents.

These results are general, applicable
for all $t$. Our interest lies in the
long time steady state limit, which
is the focus of the next subsection.

\subsection{Steady state limit}

In the $t\rightarrow \infty$ limit, we
already know that  $P(b,m;t)\to X_m^b$
(Section 2.2-3). For the other two,
our expectations of the asymptotic
behavior lead us to the ansatz:
\begin{eqnarray}
\phi (b,m) &\to &\omega _m^bt+\Phi _m^b\quad ,  
\nonumber \\
\psi (b,m) &\to &\lambda _m^bt+
\Psi _m^b\quad .  \label{harminckilenc}
\end{eqnarray}

To determine the four unknowns,
we insert these asymptotic forms into (\ref
{huszonnegy}- \ref{huszonhat}) and
obtain a systems of equations for them.
Letting $\alpha $ denote either
$\omega $ or $\lambda $, we have
\begin{eqnarray}
 {m=0:}\qquad \qquad 0=\hspace*{-0.6cm}&&
{\cal J}_1^b(\alpha )-{\cal J}_0^b(\alpha
)  \nonumber \\
 {m\neq 0,N:}\qquad \qquad 0=\hspace*{-0.6cm}&&
{\cal J}_{m+1}^b(\alpha )-{\cal J}
_m^b(\alpha )  \label{negyven} \\
 {m=N:}\qquad \qquad 0=\hspace*{-0.6cm}&&
{\cal J}_0^{b+1}(\alpha )-{\cal J}
_N^b(\alpha )  \nonumber
\end{eqnarray}
For  $\Phi _m^b$ and $\Psi _m^b$, we find
\begin{eqnarray}
 {m=0:}\qquad \qquad \omega _0^b
=\hspace*{-0.6cm}&&{\cal J}_1^b(\Phi )-{\cal J}
_0^b(\Phi )+X_N^{b-1}q  \nonumber \\
 {m\neq 0,N:}\qquad \qquad \omega_m^b
=\hspace*{-0.6cm}&&{\cal J}_{m+1}^b(\Phi )-{\cal
J}_m^b(\Phi )  \label{negyvenegy} \\
 {m=N:}\qquad \qquad
\omega _N^b=\hspace*{-0.6cm}&&
{\cal J}_0^{b+1}(\Phi )-{\cal J}
_N^b(\Phi )-X_0^{b+1}q^{\prime }  \nonumber
\end{eqnarray}
and
\begin{eqnarray}
 {m=0:}\quad \lambda _0^b
=\hspace*{-0.6cm}&&{\cal J}_1^b(\Psi )-{\cal J}_0^b(\Psi
)+X_N^{b-1}q-X_1^bW_{0,1}^b  \nonumber \\
 {m\neq 0,N:}\quad \lambda _m^b
=\hspace*{-0.6cm}&&{\cal J}_{m+1}^b(\Psi )-{\cal J}
_m^b(\Psi )+X_{m-1}^bW_{m,m-1}^b-
X_{m+1}^bW_{m,m+1}^b  \label{negyvenketto} \\
 {m=N:}\quad \lambda _N^b
=\hspace*{-0.6cm}&&{\cal J}_0^{b+1}(\Psi )-{\cal J}
_N^b(\Psi )+X_{N-1}^bW_{N,N-1}^b-
X_0^{b+1}q^{\prime }.  \nonumber
\end{eqnarray}\\
\noindent
Now, (\ref{negyven}) is the same set
of equations as $X_m^b$, so that
$\omega _m^b$ and $\lambda _m^b$
must be both propoertional to $X_m^b$. Let
us write\\
\begin{eqnarray}
\omega _m^b=\hspace*{-0.6cm}&&V_{bond}X_m^b\quad ,
\label{negyvenharom} \\
&& \nonumber \\
\lambda _m^b=\hspace*{-0.6cm}&&V_\mu X_m^b\quad ,
\label{negyvennegy}
\end{eqnarray}\\
\noindent
where the constants $V_{bond}$ and
$V_\mu $ can be fixed by the
normalization condition\\  
$\sum_{b=1}^N\sum_{m=0}^NX_m^b=1$:\\
\begin{eqnarray}
V_{bond}=\hspace*{-0.6cm}&&\sum_{b=1}^N\sum_{m=0}^N
\omega _m^b\quad ,  \label{negyvenot} \\
&& \nonumber \\
V_\mu=\hspace*{-0.6cm}&&\sum_{b=1}^N\sum_{m=0}^N
\lambda _m^b\quad .  \label{negyvenhat}
\end{eqnarray}\\
\noindent
Using (\ref{tinhut}-\ref{tunhit}) and
(\ref{harminckilenc}), we arrive at
the meaning of these constants:\\
\begin{eqnarray}
V_{bond}=\hspace*{-0.6cm}&&\lim_{t\to \infty }
\frac{d}{dt}\langle \beta(t) \rangle=\lim_{t\to
\infty }v_{bond}(t)\quad ,  \label{negyvenhet} \\
V_\mu =\hspace*{-0.6cm}&&\lim_{t\to \infty }
\frac{d}{dt}\langle \mu(t) \rangle=\lim_{t\to \infty}
v_\mu (t)\quad .  \label{negyvennyolc}
\end{eqnarray}\\
\noindent
In a similar way, we have $V_j\equiv \lim_{t\to \infty }
d(\langle v_{j(t)} \rangle )/dt=V_{bond}$,
$V_{rel}=V_\mu -V_{bond}$ and $V_{hole}=V_\mu $.

Next, we apply the results from Sect. 2.2-3, namely:\\
\begin{equation}
\lim_{t\to \infty }{\cal J}_m^b(P)=
{\cal J}_m^b(X)=c=-\frac{V}{M}\quad ,
\end{equation}\\
\noindent
and find, from (\ref{huszonhet}-\ref{huszonnyolc}),\\
\begin{eqnarray}
V_{bond}=\hspace*{-0.6cm}&&-Nc=\frac{V}{N+1}\quad ,
\label{negyvenkilenc} \\
&& \nonumber \\
V_\mu=\hspace*{-0.6cm}&&-Mc=V\quad ,  \label{otven} \\
&& \nonumber \\ 
 V_j=\hspace*{-0.6cm}&&\frac{V}{N+1}\quad ,  \label{otv} \\
&& \nonumber \\ 
V_{rel}=\hspace*{-0.6cm}&&\frac{N}{N+1}V\quad ,  \label{oo} \\
&& \nonumber \\ 
V_{hole}=\hspace*{-0.6cm}&&V\quad .  \label{ooo}
\end{eqnarray}\\
\noindent
Note that these results can also be
obtained, according to (\ref{negyvenot})
and (\ref{negyvenhat}), by summing
(\ref{negyvenegy}) and (\ref{negyvenketto}
) over $m$ and $b$.

Finally, we turn our attention to
the diffusion constants (\ref{kilenc} -
\ref{kilencvesz}). Using the linearity
property of ${\cal J}_m^b(\alpha )$
as function of $\alpha $ , as well
as the expressions (\ref{negyvenharom}-
\ref{negyvennegy}) and (\ref{negyvenhet}
-\ref{negyvennyolc}) with (\ref
{huszonhet}-\ref{huszonnyolc}), we
verify that terms linear in $t$ cancel.
The expressions are somewhat simpler if
we used the renumbering scheme (\ref
{renumber}-\ref{kabe}) for the quasi-walker.
Accordingly, all quantities of
the form $A_m^b$ are now denoted by
$\overline{A}_k$, where $k=1,..,M.$ In
this notation, the current-density
type quantities in (\ref{huszonharom})
are summarized in one line:\\
\begin{equation}
\overline{{\cal J}}_k(\overline{A})=
\overline{A}_k\overline{W}_{k-1,k}-
\overline{A}_{k-1}\overline
{W}_{k,k-1}\quad ,\;\;\;\;k=1,..,M\quad . \label{jeka}
\end{equation}\\
\noindent
As in the previous section, we can solve
these for $\overline{A}_k$. If we
sum over $k$, an important relation is obtained:\\
\begin{equation}
\sum_{k=1}^M\overline{A}_k=
\left(-\frac{\Gamma M}{V}\right) \sum_{k=1}^M\overline{
u}_k\overline{{\cal J}}_
{k+1}(\overline{A})\quad ,  \label{sumA}
\end{equation}\\
\noindent
where $\overline{u}_k$ is given by (\ref{uk}).
Using this notation, the diffusion constants are:\\
\begin{equation}
D_{bond}=-\sum_{b=1}^N{\cal J}_0^b(\Phi )+
\frac{1}{2}\sum_{b=1}^N\left(
X_0^bq^{\prime }+X_N^{b-1}q\right)
-V_{bond}\sum_{b=1}^N\sum_{m=0}^N\Phi
_m^b\quad ,  \label{otvenhat}
\end{equation}
\begin{eqnarray}
D_{hole}=-\hspace*{-0.6cm}&&\sum_{k=1}^M
\overline{{\cal J}}_k(\overline{\Phi }
)-V\sum_{k=1}^M\overline{\Phi }_k+
\sum_{b=1}^N\overline{X}_{k_b-1}q
\nonumber \\
+\hspace*{-0.6cm}&&\frac N{N+1}\left[ -\sum_{k=1}^M
\overline{{\cal J}}_k(\overline{\Psi}
)-V\sum_{k=1}^M\overline{\Psi }_k+
\sum_{k=1}^M\overline{X}_{k-1}\overline{W}
_{k,k-1}\right]   \nonumber \\
-\hspace*{-0.6cm}&&\frac V{N+1}\sum_{k=1}^Mm\overline{X}_k
-\frac{V}{2}\cdot\frac{1}{N+1}\quad .
\label{otvennyolc}
\end{eqnarray}\\
\noindent
Now, we need to find expressions
for $\overline{{\cal J}}_k(\overline{\Phi })$
and $\overline{{\cal J}}_k(\overline{\Psi })$.
These can be found by
solving (\ref{negyvenegy}) and (\ref{negyvenketto}):\\
\begin{eqnarray}
\overline{{\cal J}}_k(\overline{\Phi })
=\hspace*{-0.6cm}&&\frac{V}{N+1}\frac{1}{M}
\sum_{k^{\prime }=1}^Mk^{\prime }
\overline{X}_{k^{\prime }+k-1}+\frac{V}{N+1}
\frac{m}{M}  \nonumber \\
-\hspace*{-0.6cm}&&\frac{V}{N+1}\frac{1}{M}\sum_{k^{\prime }
=1}^Mk^{\prime}\overline{X}
_{k^{\prime }}+\delta _{k,k_b}
\overline{X}_{k_b-1}q+c_{\Phi \quad }, \label{jekafi}
\end{eqnarray}\\
\noindent
and\\
\begin{equation}
\overline{{\cal J}}_k(\overline{\Psi })
=\frac{V}{M}\sum_{k^{\prime
}=1}^Mk^{\prime }\overline{X}_{k^{\prime }
+k-1}+\overline{X}_{k-1}\overline{W
}_{k,k-1}-\frac{V}{M}\sum_{k^{\prime}=1}^M
k^{\prime }\overline{X}_{k^{\prime
}}+c_\Psi \quad ,  \label{jekapsi}
\end{equation}\\
\noindent
where $m=(k-1)$ $mod$ $(N+1)$ and
$c_{\Phi (\Psi )}$ are constants which
will disappear from the final expressions.
Using (\ref{sumA}) and substituting  (\ref
{jekafi}, \ref{jekapsi}) into
(\ref{otvenhat}, \ref{otvennyolc}), we finally
arrive at explicit forms for the diffusion constants:\\
\begin{eqnarray}
D_{bond}=\hspace*{-0.6cm}&&\frac{\Gamma^2}{(N+1)^2}
\left[ V\sum_{k=1}^M\overline{u}
_k\sum_{k^{\prime }=1}^Mk^{\prime }
\overline{r}_{k^{\prime }+k}+M\sum_{k=1}^M
\overline{u}_{k_b-1}\overline{r}_
{k_b-1}q\right]   \nonumber \\
+\hspace*{-0.6cm}&&\frac{\Gamma V}{(N+1)^2}\sum_{k=1}^M
(\overline{u}_{k-1}-\overline{r}_k)m-
\frac{M+2}{2}\frac{V}{(N+1)^2}\quad ,  \label{debond}
\end{eqnarray}\\
\noindent
and\\
\begin{eqnarray}
\hspace*{-0.6cm}&& D_{hole} = \Gamma^2\left( V\sum_{k=1}^M
\overline{u}_k\sum_{k^{\prime
}=1}^Mk^{\prime }\overline{r}_
{k^{\prime }+k}+M\sum_{k=1}^M\overline{u}_k
\overline{r}_k\overline{W}_{k+1,k}\right)
-V\frac{M+2}{2}  \nonumber \\
+\hspace*{-0.6cm}&&\Gamma^2N\sum_{k=1}^M\left[ \overline{u}_
{k_b-1}\overline{r}_{k_b-1}q-
\overline{u}_{k-1}\overline{r}_{k-1}
\overline{W}_{k,k-1}\right] +\frac{\Gamma V}{
N+1}\sum_{k=1}^M(\overline{u}_{k-1}-
\overline{r}_k)m.  
\label{dehol}
\end{eqnarray}
\\
By comparing these formulae with
that for the quasi-walker, (\ref{diffc}),
the following simple relations are found:\\
\begin{equation}
D_{bond}=\frac{1}{(N+1)^2}D+\frac{1}
{N+1}\Delta \quad ,  \label{egyszeru}
\end{equation}\\
\noindent
and\\
\begin{equation}
D_{hole}=D+\Delta \quad ,
\end{equation}\\
\noindent
where $\Delta$ is\\
\begin{equation}
\Delta =\Gamma^2N\sum_{k=1}^M\left[
\overline{u}_{k_b-1}\overline{r}_{k_b-1}q-
\overline{u}_{k-1}\overline{r}_{k-1}
\overline{W}_{k,k-1}\right]+\frac{\Gamma V}{
N+1}\sum_{k=1}^M(\overline{u}_{k-1}-
\overline{r}_k)m.  \label{deltaa}
\end{equation}
\\
Eliminating $\Delta$, the diffusive
properties of the hole, the DB and the
quasi-walker can be related:\\
\begin{equation}
D_{hole}=(N+1)D_{bond}+\frac{N}{N+1}D.  \label{vegso}
\end{equation}
\\
These relationships, together with
those for the velocities (\ref
{negyvenkilenc} - \ref{ooo}), summarize
and confirm our intuitive picture
of the asymptotic properties associated
with not only the physical objects,
the impurity (DB) and the hole, but
also the quasi-walker. In the next
section, we will analyze some special
cases in detail, in order to
illustrate the use of these results.

\section{Special cases}

In this Section we consider some 
simple, yet interesting, cases. The
explicit expressions for the velocities 
and diffusion constants will be
transparent, painting a clear picture 
of the physical situation. We will
restrict ourselves to two main 
limits: {\em i)} the pure asymmetric and
symmetric cases and {\em ii)} the 
random symmetric case. In the former,
except for jumps across the defect bond, 
the particle-hole exchange rates to
the left (right) are uniform, given by 
$W_{\leftarrow }$ ($W_{\rightarrow }$ 
). Thus, the term ``pure asymmetric 
case'' was coined, while the ``pure
symmetric case'' consists of a 
further reduction: $W_{\leftarrow
}=W_{\rightarrow }$. Across the DB, 
the exchange rates (of any particle with
the hole) are kept general: $q$ and 
$q^{\prime }$, though we also consider
the limit $q=q^{\prime}$ \cite{TZ}. 
For case {\em ii)}, the exchange rate of
any particular particle with the hole 
is randomly chosen, but indepedent of
the jump direction -- thus the term 
``random symmetric''. The random {\em  
asymmetric} case is extremely interesting, 
being the most general. However,
its treatment is quite involved and 
is beyond the scope of this paper.

As we saw in the preceeding Section, 
the velocity and diffusion constant of
the quasi-walker determines almost 
all interesting quantities of the system.
Therefore, we first give these 
expressions in the particular cases.

\subsection{Pure asymmetric 
and symmetric cases}

To be specific, for the 
asymmetric case, we have, for all $n$:
\begin{equation}
W_{n+1,n}\equiv W_{\rightarrow}
\mbox{ \quad and\quad }W_{n-1,n}\equiv
W_{\leftarrow }\quad .  \label{jobbal}
\end{equation}
Since all the hopping rates in each 
direction are identical, the particles
in the system are no longer 
distinguishable. It is meaningless to specify
``the location of the DB with 
respect to the chain.'' We can just as easily
regard the hole as a particle, 
suffering biased diffusion (if $W_{\leftarrow
}\neq W_{\rightarrow }$) everywhere 
except across a specific bond (where the
bias is $q-q^{\prime }$). The model 
reduces to a simple driven system with a
single blockage \cite{blockge}, so 
that its properties can be relatively
easily understood.

Define\\
\begin{equation}
s\equiv W_{\leftarrow }/W_
{\rightarrow }\quad ,  \label{s}
\end{equation}\\
\noindent
so that $\lambda =s^N$ (from 
(\ref{l})). Using the expressions from Sect.
2.2-3, we then find: \\
\begin{eqnarray}
\overline{r}_{k_b+m} =
\frac{U}{(W_{\rightarrow }
-W_{\leftarrow })} \Bigg\{
1-\lambda s^{-m}+\frac{\lambda s^{-m}}
{q-q^{\prime }\lambda }
\Big[ W_{\rightarrow }-W_{\leftarrow} 
&+&  q^{\prime}(1-\lambda ) \Big] \Bigg\}
\quad ,\nonumber \\
\hspace*{-0.6cm}&&\quad m \in [0,N] 
\end{eqnarray}\\
\noindent
and \\
\begin{equation}
\overline{u}_{k_b+m} =\frac{U}
{(W_{\rightarrow }-W_{\leftarrow })}
\left[1-s^{m+1}+\frac{q^{\prime }s^m}
{q-q^{\prime }\lambda } 
(1-\lambda s) \right] \!\! ,\quad 
m\in [0,N-1] \label{ukabem}
\end{equation}\\
\noindent\\
\begin{equation}
\overline{u}_{k_b+N}=\frac{U}
{q-q^{\prime}\lambda}
\cdot \frac{1-\lambda s}{1-s}
\quad ,\;\;\;m=N\quad .  \label{ukabeN}
\end{equation}\\
\noindent
where $U$ is defined by (\ref{v}).
The normalization constant now takes the form:\\
\begin{equation}
\Gamma ^{-1}=\frac{UN}
{W_{\rightarrow}-W_{\leftarrow}}
\left[ N+1+
\frac{W_{\rightarrow}-
W_{\leftarrow}+q^{\prime}-q}
{q-q^{\prime}\lambda}\cdot
\frac{1-\lambda s}{1-s}
\right] .  \label{normaliz}
\end{equation}\\
\noindent
Note that, due to the translational 
invariance of the pure case, the steady
state probability distribution 
$X_m^b$ is independent of $b$, the DB
position. Using (\ref{34}), we find 
the quasi-walker's velocity $V$ to be:\\
\begin{equation}
V=(W_{\rightarrow }-W_{\leftarrow })
\left[ 1+\frac{1}{N+1}\left( 1-\frac{ 
q-q^{\prime }}{W_{\rightarrow }-
W_{\leftarrow }}\right) \frac{W_{\rightarrow
}^{N+1}-W_{\leftarrow }^{N+1}}
{qW_{\rightarrow }^N-q^{\prime }W_{\leftarrow
}^N}\right] ^{-1}\quad .  \label{purseb}
\end{equation}
\\
Since the velocities of the physical 
objects are directly related to this $V$ 
, it is of interest to point out 
two appealing aspects of this expression.
The effect of the impurity clearly 
affects $V$ at the level of $O(1/N)$.
Further, it can be completely 
eliminated by choosing $q^{\prime
}=W_{\leftarrow }\;$and$\;q=W_{\rightarrow }$, 
where $V$ reduces to the
expected result: $W_{\rightarrow }-
W_{\leftarrow }$. In this limit, the
probability distribution in fact 
becomes uniform, i.e., $X_m^b=1/M$.

A similar expression of the diffusion 
constant $D$ can be given. It is quite
cumbersome and not very transparent. 
So, instead of providing the formula
here, let us direct the reader to the 
general case discussed in the last two
Sections. However, it is easy to check 
that, if we ``eliminate'' the DB as
above, we recover the expected results: 
$D=(W_{\rightarrow }+W_{\leftarrow
})/2$, $\Delta =0$, $D_{hole}=D$, and 
$D_{bond}=D_{hole}/(N+1)^2$. The first
of these was obtained by Derrida 
\cite{Derrida}. The last equation implies
that a referrence point (tagged 
``defect bond'') on the checkerboard
diffuses with respect to the 
string of particles at a rate $N^{-2}$ times
slower than the hole itself. 
In other words, the string of particles, as a 
{\em whole,} has a diffusion 
constant given by $D_{bond}$.

The case pointed out in Ref. 
\cite{TZ}, applicable to the reptation model of
a polymer with a single drive-insensitive 
impurity, is given by $q^{\prime
}=q\equiv B$. From (\ref{purseb}) 
we see that the impurity reduces the drift
velocity by $O(1/N)$. For this 
case, the interesting generalization of our
model lies in the study of chains 
with a fixed fraction of holes and/or
impurities.

From the ``pure asymmetric case'', 
we can simplify further to the symmetric
limit, in which $W_{\rightarrow }=
W_{\leftarrow }\equiv W$ with arbitrary $q$
and $q^{\prime }$. The corresponding 
physical picture is a hole (or a
particle) performing simple random 
walk everywhere except across one bond,
where it receives an extra ``kick'' 
of strength $q-q^{\prime }$ to the
right. Thus, this case can be 
mapped into simple diffusion with open
boundaries, coupled to chemical 
potentials. Solving the ordinary diffusion
equation with unequal densities at 
the two boundaries is trivial: the result
is a linear profile. Indeed, this 
is reproduced here. Setting $s=\lambda =1$ 
, we find the probability distribution: \\
\begin{equation}
X_m^b=\frac{1}M\frac{q(N-m)+q^{\prime }m+W}
{\overline{q}N+W},  \label{ixbemw}
\end{equation}\\
where $\overline{q}=(q+q^{\prime })/2$ 
is the average jump rate through the
defect bond. The velocity, which is 
related to the current, also takes a
simple form: \\
\begin{equation}
V=\frac{q-q^{\prime }}{\displaystyle 1+
N\frac{\overline{q}}W} \quad .
\label{szepseb}
\end{equation}\\
\noindent
However the expression for the 
diffusion constants remain rather involved.
They only simplify with the further 
restriction: $q=q^{\prime }$ (where
clearly the probability distribution 
is uniform and $V=0$). Then \\
\begin{equation}
D=q\frac{1+N}
{\displaystyle 1+N\frac{q}{W}}\quad,  \label{szep}
\end{equation}\\
\noindent
with $\Delta =0$, $D_{hole}=D$, and $D_{bond}=D/(N+1)^2$.

\subsection{Random symmetric case}

In this part we study the case 
where the $W$ jump rates are symmetric: \\
\begin{equation}
W_{n,j}=W_{j,n}  \label{symm}
\end{equation}\\
\noindent
but otherwise randomly chosen out 
of some distribution. The rates $q$ and $ 
q^{\prime }$ are also kept in general.

An immediate consequence is 
$\lambda =1$. Using the expressions 
(\ref{70}-\ref{62}) for $\overline{r}_k$ 
and $\overline{u}_k$ and introducing \\
\begin{equation}
z\equiv \sum_{l=1}^N\frac{1}{W_{l+1,l}}  \label{ze}
\end{equation}\\
\noindent
for convenience, we arrive at: \\
\begin{equation}
\overline{r}_{k_b+m}=\left[1-
\left( \frac{q^{\prime}}q\right) ^N \right]
\left[\frac{1+q^{\prime}z}{q-q^{\prime }}+
\sum_{i=1}^{N-m}\frac{1}{ 
W_{b+m+i,b+m+i-1}}\right]\;\;,\;\;m\in [0,N]  \label{erm1}
\end{equation}\\
\noindent
and \\
\begin{equation}
\overline{u}_{k_b+N}=\frac{N+1}{q-q^{\prime }}
\left[1-\left(\frac{q^{\prime }}q\right)^N\right]
\quad, \label{ur1}
\end{equation}\\
\noindent\\
\begin{equation}
\overline{u}_{k_b+m}=\frac{1}{W_{b+m+1,b+m}}
\left[ 1-\left( \frac{q^{\prime }}q 
\right) ^N\right]
\frac{q(m+1)+q^{\prime }(N-m)} 
{q-q^{\prime }} \quad .
\label{urm1}
\end{equation}\\
\noindent
From the normalization constant\\ 
\begin{equation}
\Gamma ^{-1}=M\left[ 1-\left( 
\frac{q^{\prime }}q\right)^N\right]
\frac{1+\overline{q}z} 
{q-q^{\prime }}  
\quad ,
\end{equation}\\
\noindent
we obtain the velocity: \\
\begin{equation}
V=\frac{q-q^{\prime }}{1+
\overline{q}z}\quad .  \label{joseb}
\end{equation}\\
\noindent
This formulae also displays some 
intuitively understandable properties,
e.g., proportionality to the drive 
$(q-q^{\prime })$ and being of $O(1/N)$
(through $z \thicklines \sim \thinlines O(N)$). As in 
the previous cases, explicit expressions
for the diffusion constants are not 
particularly illuminating and will not
be given here. However, in the 
{\em totally} symmetric case, i.e., $ 
q=q^{\prime }$, there is simplification:\\ 
\begin{equation}
D=q\frac{1+N}{1+zq}  \label{utolso}
\end{equation}\\
\noindent
with $\Delta =0$, $D_{hole}=D$, 
and $D_{bond}=D/(N+1)^2$.

If the $W$'s are drawn from the distribution $\rho (W)$, then the
expressions for the velocity and diffusion constants are valid in the
thermodynamic limit $N \to \infty$ with $z/N=\int \rho (W)\frac{dW}W$.
Finally, we point out the obvious: for a $\delta$ 
distribution, (\ref{joseb},
\ref{utolso}) reduce to (\ref{szepseb},\ref{szep}).

\section{Summary and outlook}

In this paper, we study a generalized 
asymmetric exclusion process (ASEP) 
\cite{Derrida} to include a 
directional impurity. The simplest realization
consists of a random walker 
(a hole) on a one dimensional periodic lattice
which is otherwise filled with 
particles. The rates at which the hole
exchanges with any particular 
particle $i$ ($i=1,...,N$) is arbitrary and
direction-dependent (but time-indepedent). 
In otherwords, the rate of a
particle hop to the right, 
$W_{i+1,i}$, can be different from those of a hop
to the left, $W_{i-1,i}$. However, 
across one special bond (which is fixed in respect
to the {\em lattice} and not the string of particles), 
which we call the ``defect bond'', the rates of 
particle-hole exchange is fixed at $q$ and 
$q^{\prime }$, regardless of which 
particle is involved. In general, the $q$ 
's are of course unrelated to the 
$W$'s. In this sense, there is an
``impurity'' in the lattice, associated 
with the mobility of the walker,
which breaks translational invariance.

Though the problem appears to have 
two degrees of freedom -- the positions
of the hole and the defect bond -- we 
show that the associated configuration
space is, in fact, one-dimensional. 
There is a subtle coupling of these two
objects into a new entity which we 
label the quasi-walker. To be precise,
the location of the bond, $b$, and the 
bond-hole separation, $m$, can be
combined into a single variable $k\equiv 
1+(b-1)(N+1)+m$, running from $1$
to $M\equiv N(N+1)$. A Master equation 
for $\overline{P}_k(t)$, the
probability to find the system in the 
state $k$ at time $t$, can be easily
written. Being a one-dimensional 
problem, it can be solved and the steady
state distribution found explicitly. 
Further, generalizing a 
method used in the theory of random
walks on lattice rings \cite{Hughes}, \cite{Derrida}, 
(replication) we 
are able to find implicitly
time-dependent information in the 
steady state, namely, asymptotic
velocities and diffusion constants 
associated with both the hole and the
defect bond. The final expressions, 
though explicit functions of the $q$'s
and $W$'s, are rather complicated, 
so that it is difficult to gain much
insight into their physical content. 
Thus, we considered a few special
cases, for which these explicit 
formulae display intuitively comprehensible
properties of the system, such as 
the order of magnitude of the effects of
the impurity.

Turning our focus to physical 
realizations of this model, we rely on the
existing connection between an 
impurity-free ASEP and the Duke-Rubinstein
reptation model for gel-electrophoresis 
\cite{rubi}. There, a polymer
is thought of as a chain of segments, 
confined to a tube of cells. Each cell
may contain one or more segments. 
The {\em links} between segments are
mapped into particles and/or holes in 
the ASEP in that a link bridging two
cells corresponds to a particle encoded 
with ``charge'' $\pm 1$, while a
link lying within one cell corresponds 
to a hole. In an external field, the
segments move preferentially in the 
direction of the field, a behavior which
translates into the ``pure asymmetric 
case'' of the ASEP. Of course, in
general, a polymer will have more than 
one link lying within a cell,
motivating the generalization of 
Derrida's work to a multi-hole ASEP
\cite{KvL}. A more fundamental difference 
between a periodic ASEP and
gel-electrophoresis lies in that 
open boundary conditions are more
appropriate for the latter, leading 
to many changes in the string of
``particles'' in the ASEP. Significantly, 
a recent work \cite{P-Spohn}
proved rigorously that, to leading order 
in $N$, the diffusion constant of
an open chain coincides with that of 
the periodic chain. Thus, the analysis
of a periodic chain \cite{Derrida,KvL} 
should be taken more seriously than
an academic exercise. In a previous 
publication \cite{TZ}, we have shown
that a segment of the polymer which 
is insensitive to the external drive (an
impurity) can be mapped into a defect 
bond in the ASEP. Based on the success
of the impurity-free ASEP as a model 
for gel-electrophoresis, the work here
should be further generalized to 
include multi-hole and multi-impurity
cases. The results may be of importance 
to experiments on the drifting of
polymers made from a mixture of 
two different types of monomers.

Other applications of asymmetric exclusion 
processes include, e.g., driven
diffusive systems with two species 
\cite{shz} and models of traffic flow 
\cite{traffic}. The interesting 
phenomena of phase transitions observed in
these systems are, however, present 
only in two or more dimensions. It
clearly behooves us to examine our 
model in dimensions higher than one.
Another obvious extension of our 
study is the effect of open boundary
conditions. The impurity-free ASEP, 
even in one-dimension, is known to
display extremely interesting collective 
behavior such as spontaneous
symmetry breaking \cite{asepobc}. 
There is little doubt that our study
should be regarded as only a small 
step towards the exploration of the vast
unknown of non-equilibrium statistical mechanics.

\subsection*{Acknowdlegements}

Illuminating discussions with M. Evans,
C. Hill, C. Laberge, G. Korniss, Y. Shnidman, G.
Sch\"{u}tz, T. T\'{e}l, and
especially B. Schmittmann are gratefully
acknowledged. We are indebted to 
T.T\'{e}l and Z.R\'{a}cz for the kind
hospitality and support at the 
E\"{o}tv\"{o}s Workshop, Budapest, July 1994,
where this work was first presented.
One of us (RKPZ) thanks E. Domany and
D. Mukamel for their hospitality at
the Weizmann Institute, Israel. This
research is supported in part by the
US National Science Foundation through
the Division of Material Research and
the Hungarian Science Foundation under
grant number OTKA F17166.

\end{document}